# Generative AI for Real-Time Ultrafast Pulse Characterization from Sparse Measurements

**Abhimanyu Borthakur,**[a,*] **Jack Hirschman,**[b,c] **Sergio Carbajo**[a,c,d,e]

[a]Department of Electrical & Computer Engineering, University of California Los Angeles, Los Angeles, CA 90095, USA

[b]Department of Applied Physics, Stanford University, Stanford, CA 94305, USA

[c]SLAC National Accelerator Laboratory, Stanford University, Menlo Park, CA 94025, USA

[d]Department of Physics and Astronomy, University of California, Los Angeles, CA 90095, USA

[e]California NanoSystems Institute, 570 Westwood Plaza, Los Angeles, CA 90095, USA

**Abstract**. Spectrographs and spectrometers are foundational instruments across optics and photonics, enabling applications from compact spectral imaging and chip-scale spectroscopy to field and space remote sensing. Yet they routinely operate under size, speed, and photon-budget constraints that yield sparse, noisy, or incomplete spectral data. These same constraints are increasingly central to ultrafast pulse metrology—where the information of interest is often encoded in spectrogram-like measurements, whose acquisition demands dense scans across coupled spectro-temporal axes. In critical, resource-constrained or high-speed operational environments—such as industrial laser machining, biomedical imaging, or field-deployed optical systems—dense spectro-temporal scans are often impractical. Moreover, existing computational methods falter with severely undersampled data, leading to blurred, unstable, or computationally expensive reconstructions. Here, we introduce a generative artificial intelligence framework based on implicit diffusion models that robustly retrieves ultrafast pulse intensity and phase from aggressively downsampled spectrograms. Our approach not only surpasses state-of-the-art deep learning baselines in accuracy and stability (with an order of magnitude improvement in mean absolute error) but also operates at near-real-time speeds (~52 ms per pulse), enabling deployment in online diagnostic and control systems. By transforming sparse measurements into precise pulse profiles, this work democratizes access to high-fidelity ultrafast metrology and spectrogram assimilation, extending its reach from advanced laboratories to educational, industrial, and field settings.

***First Author, E-mail: abhimanyu911@ucla.edu**

## 1 Introduction

Spectrographs and spectrometers sit at the center of modern photonics: by mapping wavelength into a measurable coordinate, they enable material identification, spectral imaging, and optical-system diagnostics across laboratory, industrial, biomedical, and remote deployments. In their many forms,[1] such as diffraction-grating spectrographs (including Czerny–Turner geometries), interferometric designs (Fourier-transform spectroscopy), filter- or etalon-based spectrometers (acousto-optic tunable filters and Fabry–Pérot implementations), and emerging integrated/reconstructive spectrometers that explicitly trade hardware complexity for computation, these instruments share a common tension: spectral resolution, throughput, acquisition time, and size/weight/power cannot be simultaneously optimized, so measurements are often sparse or incomplete when scenes are dynamic, photon-limited, or system-constrained.[2–8] Among the most stringent regimes are snapshot and compressive spectral imaging—where a full spatial–spectral datacube must be inferred from limited measurements—and field/spaceborne spectroscopy, where payload and operational constraints sharply limit acquisition time and system complexity.[5,8,9]

The advent of ultrashort pulses has revived this issue in a particularly demanding form—to be able to measure and control ultrashort laser pulses is not merely a laboratory curiosity—it is a foundational capability for modern photonics innovation. From enabling non-invasive medical diagnostics and high-precision manufacturing to powering ultrafast optical communications and probing quantum dynamics, these pulses are tools of transformative potential.[10] They enable the study of ultrafast phenomena in physics, chemistry, and biology, providing unprecedented temporal resolution down to the attosecond scale.[11,12] The development of high-intensity, ultrashort laser pulses has been recognized with multiple Nobel Prizes, including the 2018 award for their chirped pulse amplification and the 2023 award for their contributions to extreme nonlinear optics and attophysics. Modern applications of ultrashort pulses span various fields, including precision machining, microscopy, medical imaging, and telecommunications.[13–16] Recent developments in spatial light modulation have further expanded the capabilities of ultrashort pulses, enabling tailored beam profiles for novel light-matter interactions.

The complete temporal characterization of ultrafast pulses requires recovering both intensity and phase from spectro-temporal measurements that are intrinsically high-dimensional and sensitive to sampling, noise, and algorithmic convergence.[17–21] However, the gap between research-grade characterization systems and the needs of real-world applications remains wide. Many stakeholders—including university teaching labs, industrial R&D teams, and field engineers—lack access to high-resolution delay scanners or broadband spectrometers, especially in challenging spectral regimes like the ultraviolet (UV). This accessibility gap limits the adoption of ultrafast technologies beyond specialized labs—There have been many techniques developed to try and characterize ultrashort pulses.[22] Among these innovations, Frequency-Resolved Optical Gating (FROG) emerged as a groundbreaking solution.[23]

It accomplishes this through a self-referenced scheme in which the pulse is split into two replicas that overlap in a nonlinear medium, where one replica acts as a gate for the other. The spectrum of the nonlinear signal generated by their interaction is recorded while the relative delay between the replicas is scanned. The resulting two-dimensional map of intensity versus frequency and delay is effectively a spectrogram of the pulse and is known as the FROG trace.[24]

However, for online operation or control systems, time delay scans can be prohibitively expensive for long pulses (therefore, creating a bottleneck for data collection) or limited by spectrometer resolution.[25] Even if high-resolution spectrometers are deployed, they might not be appropriate for a particular wavelength of light, such as UV, where short bandwidth pulses yield only a few sample points.[26,27] This motivates the need for a new modeling framework that can democratize access to FROG technologies for stakeholders limited by the aforementioned constraints. Automated pulse reconstruction from incomplete (or partial) FROG traces is therefore an increasingly valuable pursuit. Here, 'incomplete' designates any FROG trace in which the two-dimensional delay–frequency grid is extensively undersampled, for example, retrieving a 512-point pulse array from an aggressively downsampled 64 x 64 two-dimensional trace, instead of the full 512 x 512 trace.

Traditional iterative FROG reconstruction is computationally intensive, limiting its use in large-scale simulations. Recently, convolutional neural networks (CNNs) and sequence-to-sequence (Seq2Seq) models have shown great promise in retrieving ultrafast pulse intensity and phase from FROG spectrograms, achieving much faster reconstructions than traditional iterative algorithms.[25,28] However, these deep learning methods struggle with downsampled FROG traces: CNNs cannot effectively model temporal context, and Seq2Seq models struggle with incomplete data.[29–31] Specifically, CNNs aggregate information through stacked local filters, which can lead to blurring or averaging of fine input details.[32] Further, CNNs that excel on dense measurements misfire on sparse ones.[33] Seq2Seq or sequential models, by contrast, process their inputs sequentially and therefore assume a continuous input stream. Discontinuities such as those between the pixels in partial FROG traces, hence, break the hidden-state continuity and allow vanishing-gradient or exploding-gradient pathologies to dominate, leading to unstable optimization updates.[34]

We present a generative diffusion framework that not only addresses the technical limitations of prior methods but also aligns with the practical demands of real-world photonics systems. By enabling accurate pulse retrieval from heavily undersampled FROG traces, our model bridges the gap between high-end metrology and resource-constrained environments. This work positions generative AI as a key enabler for the next generation of portable, affordable, and real-time ultrafast diagnostic tools. We compare our diffusion approach with other techniques by benchmarking them on a simulated dataset of FROG-pulse pairs. While this study is broadly applicable to FROGs and other spectrogram-based techniques, we focus our attention on Second Harmonic Generation (SHG) FROGs for brevity.

## 2 Methods

### *2.1 Data generation*

We generate synthetic data by simulating sinc-shaped ultrashort pulses modulated by linear chirp, self-phase modulation (SPM), and third-order phase terms. Each pulse is a 256-point array over a 100 fs window, modeled using Eq. 1, with the amplitude A(t) given by Eq. 2 and the phase $\phi(t)$ represented by Eq. 3. This E(t) is our ground truth pulse and the full SHG FROG trace is computed as the squared magnitude of the Fourier transform of the product $E(t)E(t-\tau)$, and downsampled 8x (on both axes) to produce a $32 \times 32$ spectrogram $I(\omega, \tau)$. We generate 22,000 samples, varying $t_{FWHM} \in [20, 29]$ fs, $a_{chirp} \in [0.01, 0.02]$ fs$^2$, $b_{SPM} \in [5,8.9]$ W$^{-1}$, and $c_3 \in [10^{-4}, 4 \times 10^{-4}]$ fs$^{-3}$. The dataset is split into 18,000 training, 2,000 validation, and 2,000 test tuples, each of the form (I(ω, t), (A(t), ϕ(t))), with traces normalized to unit peak intensity through division by the maximum value.

$$E(t) \ = \ A(t)\, exp[i\phi(t)], \tag{1}$$

$$A(t) \ = \ |sinc(t/t_{FWHM})|, \tag{2}$$

$$\phi(t) \ = \ a_{chirp}t^2 + b_{SPM}|A(t)|^2 + c_3t^3, \tag{3}$$

*2.2 Machine Learning (ML) Modeling and Pulse Reconstruction*

The goal of our trained ML models is to recover A(t) and Φ(t) from an unseen (test set) I(ω, τ) as input. We compare three architectures for SHG-FROG pulse reconstruction: a Visual Geometry Group (VGG) CNN, a Seq2Seq model with attention, and a diffusion model. The VGG network[35] serves as a CNN baseline due to its simplicity and strong performance in image-based tasks. The Seq2Seq model[25] demonstrated state-of-the-art accuracy on partial SHG-FROG traces, outperforming classical iterative algorithms.[36,37] Our diffusion model integrates a CNN encoder with a sequence model to capture both spatial and temporal structure and is trained using the Denoising Diffusion Implicit Model (DDIM) framework.[38] A snapshot of the three models is provided in Figure 1. For the sake of completeness, we also benchmark the performance on other models, including CapsNet,[39] ResNet,[40], and DenseNet.[41] All models are implemented in PyTorch. For this paradigm, one might be tempted to use unsupervised iterative algorithms in order to avoid data generation and training overhead; however, even fast and general-purpose solvers such as COPRA remain iteration-limited in throughput and encounter convergence issues owing to fixed step-size.[42,43]

*2.3 Diffusion model specifics*

Our diffusion model is trained using Algorithm 1. It uses two neural networks $f_{\mathrm{w}}$ (for amplitude) and $g_{\mathrm{w}}$ (for phase), which share a common encoder (see Figure 1) for the input spectrogram $I$. Each training iteration begins by sampling a diffusion timestep $t$ using a learned sampler network $\mathrm{Sampler}_\theta(I)$. This sampler outputs a categorical distribution over $t$ (of total $T$ diffusion steps), and a Gumbel-Softmax method is used to sample $t$ in a differentiable way.[44] By learning the sampling policy for $t$, the training focuses on more informative noise levels instead of uniformly sampling timesteps, similar in spirit to recent non-uniform timestep scheduling strategies that accelerate diffusion model training.[45] Given the chosen $t$, the algorithm diffuses forward: it adds Gaussian noise $\epsilon \sim \mathcal{N}(0, I)$ to the ground-truth amplitude $A$ to produce a noisy amplitude $x_t = \sqrt{\alpha_t} A + \sqrt{1 - \alpha_t}\epsilon$. Here $\alpha_t = \prod_{\tau=1}^{t}(1 - \beta_\tau)$ corresponds to a standard forward diffusion process with variance schedule $\beta_t$ – in this case a cosine noise schedule that smoothly varies noise intensity across $T$=1000 steps.[46] The amplitude network $f_{\mathrm{w}}$ then tries to denoise this sample: it takes as input the noisy $x_t$, the spectrogram $I$, and the time index $t$ (passed through a sinusoidal time embedding), and predicts $\hat{x}_0$ an estimate of the original clean amplitude. Subsequently, the phase network $g_{\mathrm{w}}$ aims to predict the spectral phase: it uses the same encoded features of $I$ along with the amplitude information to output $\hat{\Phi}$, an estimate of the true phase $\Phi$. During training $g_{\mathrm{w}}$ is given the ground-truth $A$ as input, whereas at test time, it would use $f_{\mathrm{w}}$'s predicted amplitude. The two networks are trained in succession, and the Adam optimizer[47] is used to update their parameters as well as the weights of the sampler network. The training loop repeats for 100 epochs, and throughout training, the learning rates are gradually annealed with a cosine schedule[48] from a starting value of $10^{-4}$.

Importantly, both $f_{\mathrm{w}}$ and $g_{\mathrm{w}}$ rely on the same encoder architecture to extract a 1024-dimensional feature representation $z$ from the input spectrogram $I$. Specifically, this shared encoder first applies a standard residual convolutional network backbone to process the 32×32 FROG trace into spatial feature maps. These feature maps (of shape 4×4×128 after the ResNet layers) are then flattened into a sequence of length 16 and fed into a bidirectional Long Short-Term Memory (LSTM) network to capture global structure across the image.[49] Finally, an additive attention mechanism is next applied wherein the model learns to weight and sum the LSTM outputs into a single context vector (1024-d) that represents the salient information in the spectrogram. This context vector is concatenated with the noisy amplitude $x_t$ and the time embedding before being passed through $f_{\mathrm{w}}$'s fully connected layers to predict $\hat{x}_0$, while for phase prediction, the context vector is concatenated with the amplitude vector and fed into $g_{\mathrm{w}}$'s MLP to produce $\hat{\Phi}$.

This encoder architecture is motivated by the need to jointly capture rich local spectral features and global temporal dependencies in structured time–frequency representations. Deep residual convolutional networks have been shown to learn hierarchical and discriminative spectro-temporal features from spectrogram inputs more effectively than shallow CNNs, improving both representation quality and trainability through identity skip connections.[50]

However, purely convolutional models are limited in modeling long-range temporal dependencies, particularly in spectrogram representations where meaningful structure often spans many frequency and time bins. Hybrid architectures that combine convolutional front-ends with recurrent layers such as bidirectional LSTMs have demonstrated superior performance in capturing such dependencies in tasks including speech separation, sound classification, and spectrogram regression when compared to pure CNN or pure RNN models.[51,52] Furthermore, attention mechanisms applied after recurrent processing have been empirically shown to improve the extraction of task-relevant global features by weighting time–frequency frames according to their contribution to the downstream objective, leading to performance gains in structured audio tasks.[51] Together, these results support the use of a residual CNN backbone followed by bidirectional recurrent modeling and attention to form a context vector that effectively summarizes both local and global structure in the spectrogram, which is critical for accurate amplitude and phase estimation.

Our inference procedure in Algorithm 2 (see Figure 2) follows the DDIM sampling[38] algorithm where the $S$ timesteps are sampled through our learned sampler. For our experiments, we set $S = 10$. The inference algorithm is illustrated in Figure 2.

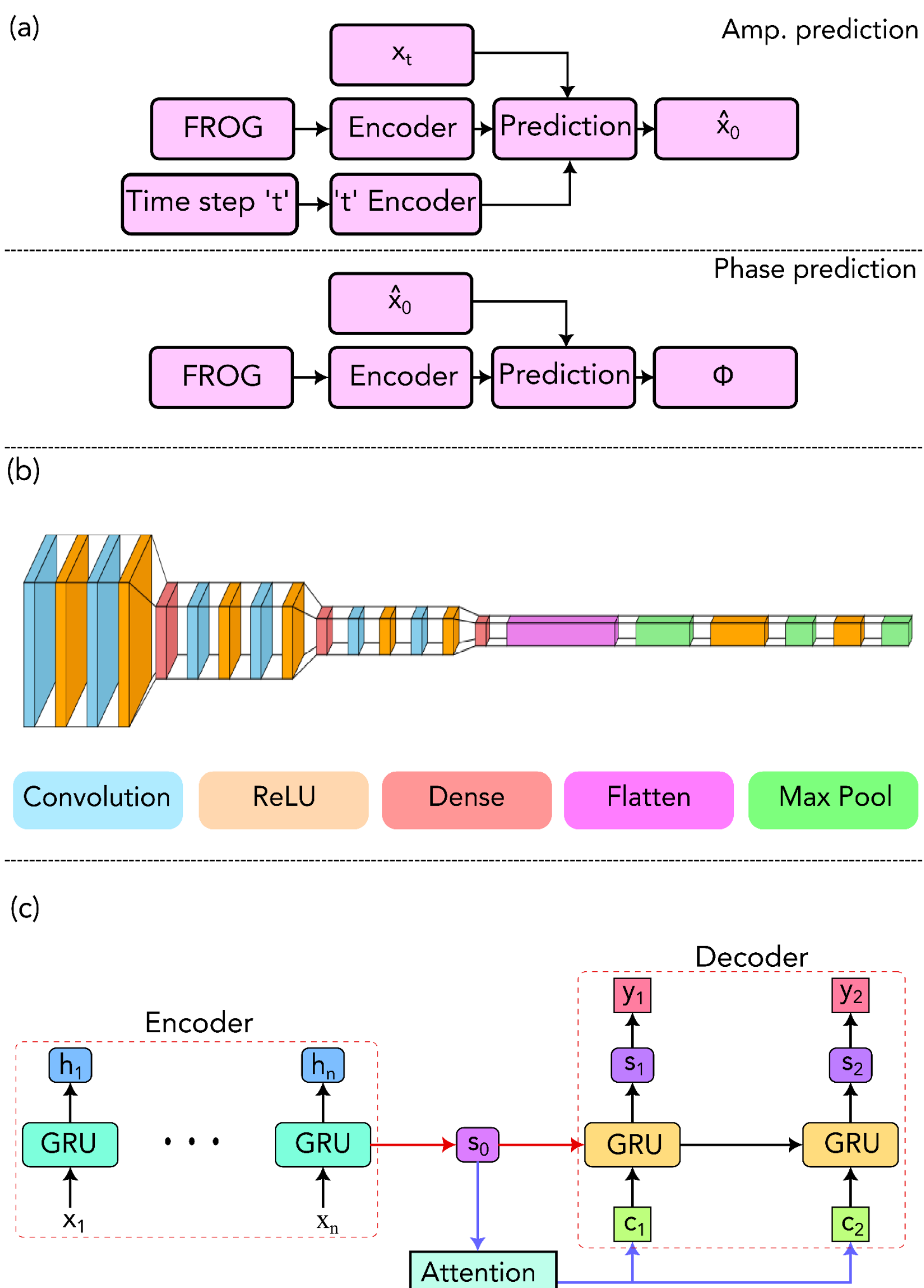

**Fig. 1** (a) Diffusion denoiser (b) VGG CNN (c) Seq2Seq network.

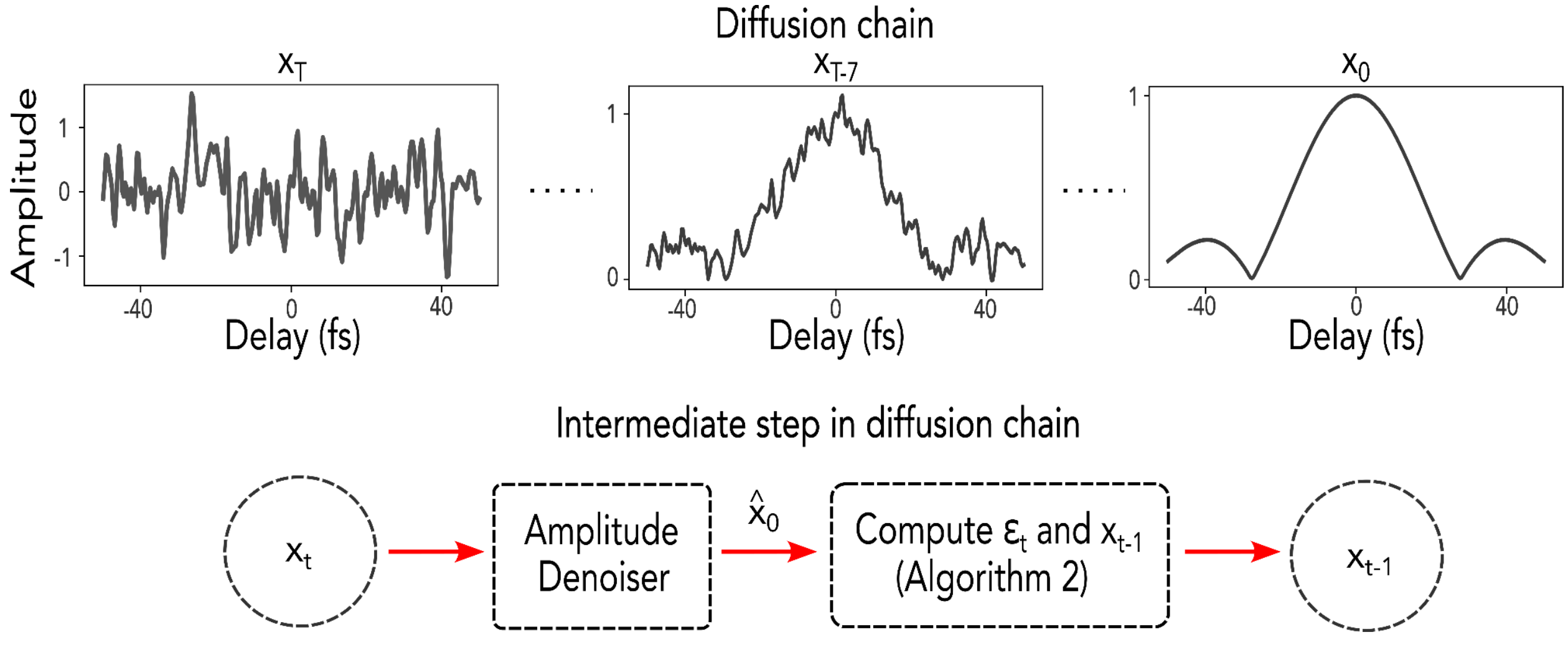

**Fig. 2** High-level visualization of Algorithm 2.

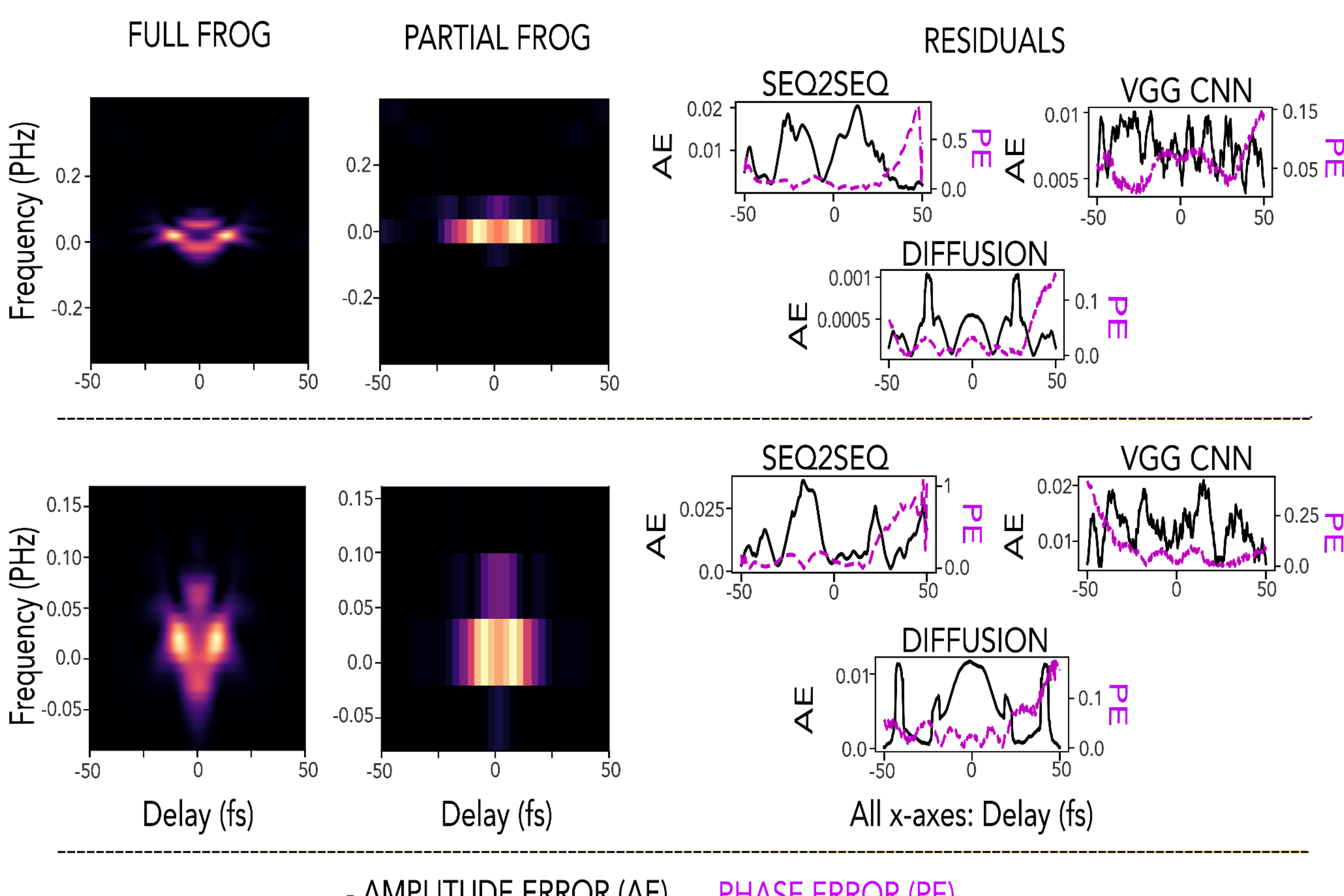

**Fig. 3** Pulse reconstruction residuals for VGG, Seq2Seq and diffusion on FROG traces sampled from 25th (top) and 50th (bottom) percentiles of the error distribution (on test set). Individual sample reconstruction errors for each model are listed in Table 2.

**Algorithm 1:** Training the Implicit Diffusion Model

**Input:** Training set $\mathcal{D} = \{(I_n, A_n, \Phi_n)\}$, noise schedule $\{\beta_t\}_{t=1}^{T}$, epochs $E$, batch size $B$, learning rates $\eta_{\text{amp}}, \eta_{\text{ph}}, \eta_{\text{sam}}$

**Output:** Model parameters $\mathbf{w}$, sampler parameters $\theta$

**for** $t \leftarrow 1$ **to** $T$ **do**

  $\alpha_t \leftarrow \prod_{\tau=1}^{t}(1 - \beta_\tau)$;

**repeat**

  Shuffle $\mathcal{D}$ into batches of size $B$;

  **foreach** *batch* $(I, A, \Phi)$ **do**

    *// 1) sample timestep via learned sampler*

    $\ell \leftarrow \text{Sampler}_\theta(I)$;

    $t \leftarrow \arg\max \text{GumbelSoftmax}(\ell)$;

    *// 2) diffuse forward*

    $\epsilon \sim \mathcal{N}(0, I)$;

    $x_t \leftarrow \sqrt{\alpha_t}\, A + \sqrt{1 - \alpha_t}\, \epsilon$;

    *// 3) predict and compute losses*

    $\hat{x}_0 \leftarrow f_{\mathbf{w}}(x_t,\, I,\, t)$;

    $\mathcal{L}_A \leftarrow \|\hat{x}_0 - A\|_1$;

    $\hat{\Phi} \leftarrow g_{\mathbf{w}}(I, A)$;

    $\mathcal{L}_P \leftarrow \|\hat{\Phi} - \Phi\|_1$;

    *// 4) update model & sampler*

    $\{\mathbf{w}, \theta\} \leftarrow \text{Adam}(\mathcal{L}_A, \mathcal{L}_P\, ;\, \eta_{\text{amp}}, \eta_{\text{ph}}, \eta_{\text{sam}})$;

**until** *epoch* $> E$ *or convergence*;

**return** $\mathbf{w}, \theta$

**Algorithm 2:** Inference

**Input:** Test trace $I$, trained models $g_{\mathbf{w}}, f_{\mathbf{w}}$, sampler $\theta$, substeps $S$

**Output:** Reconstruction $(\hat{A}, \hat{\Phi})$

$x \sim \mathcal{N}(0, I)$;

$T \leftarrow \texttt{diff_steps} - 1$;

$\ell \leftarrow \text{Sampler}_\theta(I)$;

$\{t_i\}_{i=2}^{S} \leftarrow \arg\max^{S-1}\big(\text{GumbelSoftmax}(\ell)\big)$;

$t_1 \leftarrow T$;  sort $\{t_i\}_{i=1}^{S}$ in descending order;

**for** $i \leftarrow 1$ **to** $S$ **do**

  $\hat{A}_0 \leftarrow f_{\mathbf{w}}(x,\, I,\, t_i)$;

  $\epsilon \leftarrow \big(x - \sqrt{\alpha_{t_i}}\, \hat{A}_0\big) \,/\, \sqrt{1 - \alpha_{t_i}}$;

  **if** $i < S$ **then**

    $x \leftarrow \sqrt{\alpha_{t_{i+1}}}\, \hat{A}_0 + \sqrt{1 - \alpha_{t_{i+1}}}\, \epsilon$;

  **else**

    $x \leftarrow \hat{A}_0$ ;  *// final refinement*

$\hat{A} \leftarrow x$;

$\hat{\Phi} \leftarrow g_{\mathbf{w}}(I, \hat{A})$;

**return** $\hat{A}$, $\hat{\Phi}$;

## 3 Results and Discussion

From Table 1, it is clear that the diffusion model outperforms every other model in terms of average amplitude and phase reconstruction error. The VGG and the Seq2Seq models are the second- and third-best-performing models, respectively, in terms of amplitude mean absolute error. We can observe the reconstructions of each of the top 3 models on FROG instances randomly sampled from across their error distributions in Figure 3. The amplitude reconstruction of our VGG CNN is noisy and consists of local artifacts because the strictly local convolutional receptive fields overestimate pixel-level variations.

**Table 1** Mean Absolute Error (MAE) for each model averaged on the test set

| Model | Amp MAE | Phase MAE |
|---|---|---|
| Seq2Seq[10] | 0.01256 | 0.29125 |
| VGG[20] | 0.00941 | 0.10170 |
| Diffusion (this work) | **0.00070** | **0.05375** |
| CapsNet[24] | 0.01816 | 0.13267 |
| ResNet[25] | 0.03361 | 0.20523 |
| DenseNet[26] | 0.02405 | 0.18485 |

**Table 2** Reconstruction errors for illustrative examples in Figure 3

| Percentile index | Seq2Seq | | VGG | | Diffusion | |
|---|---|---|---|---|---|---|
| | Amp MAE | Phase MAE | Amp MAE | Phase MAE | Amp MAE | Phase MAE |
| 25 | 0.0089 | 0.1412 | 0.0075 | 0.0585 | **0.0004** | **0.0304** |
| 50 | 0.0130 | 0.2429 | 0.0125 | 0.0847 | **0.0006** | **0.0458** |

The amplitude reconstruction of the Seq2Seq, however, is smoother but less accurate as the recurrent attention operation averages information across neighbouring time-steps, suppressing noise while introducing a smoothing bias. While both CNN and Seq2Seq produce visually appealing results for the phase, it is clear, from Figure 3 and Table 1, that the diffusion does best, both visually and on average. The mean inference (prediction) time of the diffusion model is 0.052 seconds per sample (batch size = 1) on a Google Colab A100 GPU.

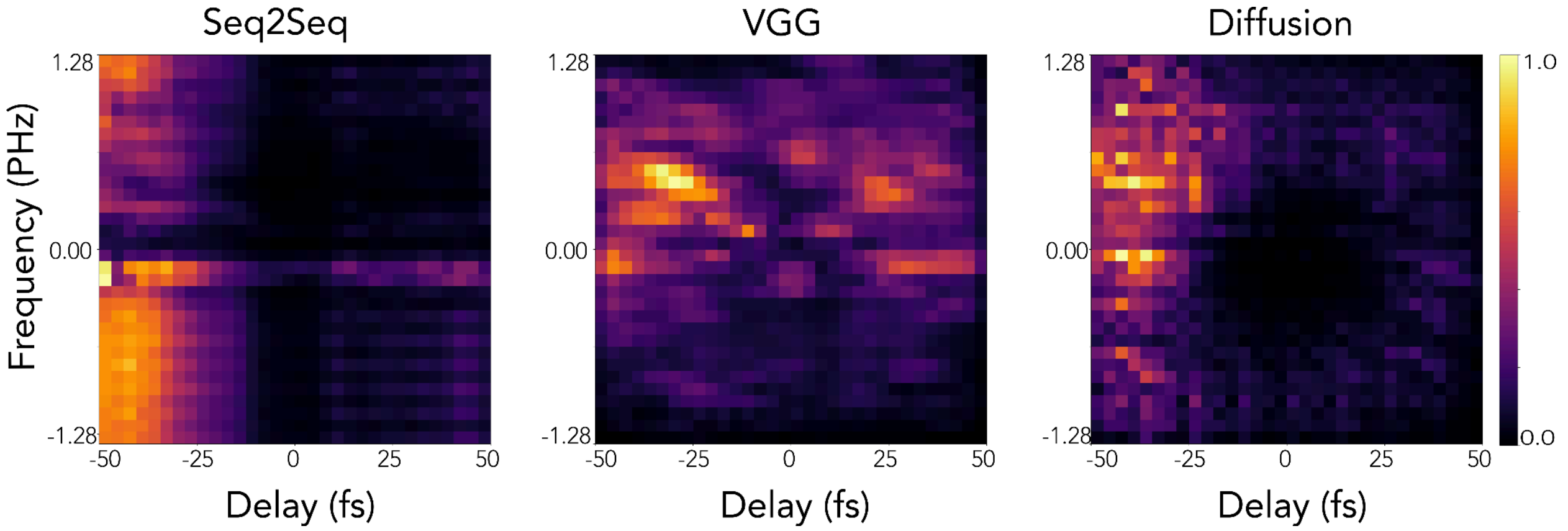

**Fig. 4** Average saliency maps obtained on the test set

The aforementioned behaviour of each model can be analysed quantitatively and qualitatively from the average saliency maps (see Figure 4) obtained on the unseen test split. These maps are built as follows: for every test set trace, we back-propagate the mean absolute error on both amplitude and phase, take the absolute input gradients of this error with respect to the 2D trace, normalize the two maps separately, fuse them with an equal-weight summation, and finally average the result over the entire test set. A bright response at any region of the map indicates that the respective network is making strong use of that region for predictions and vice versa.

The diffusion model exhibits an almost radially symmetric, low-contrast pattern with a broad annulus before tapering into subtle high-frequency streaks. This confirms that the diffusion network first distributes responsibility evenly across the trace, capturing the global pulse shape before sharpening its attention on the highest-frequency fringes to polish residual error. The convolutional VGG regressor, in contrast, lights up only a few bright blotches scattered around mid-band delays and frequencies. Since it fixates on just these local motifs and largely ignores the rest of the grid, its reconstructions end up patchy and noisy wherever those cues are missing. The Seq-to-Seq with Bahdanau attention concentrates almost all gradient mass along a single vertical strip of delays and selective horizontal slices in frequency. The recurrent attention, therefore, latches onto a few selective anchor points, propagates information temporally, and consequently outputs a smoother, yet biased reconstruction.

## 4 Conclusion

### 4.1 Enabling Accessible Ultrafast Metrology

Our diffusion-based retrieval framework significantly lowers the barriers to high-quality pulse characterization. By reducing the required sampling density by 8× along each axis, the method allows the use of lower-cost, lower-resolution hardware—such as compact spectrometers or coarse delay stages—without compromising reconstruction fidelity. This opens avenues for educational kits, field-deployable diagnostic units, and integrated quality-control sensors in manufacturing lines.[53]

### 4.2 Real-Time Control and Adaptive Systems

With inference times of ~52 ms per pulse on consumer-grade GPU hardware, our model supports real-time feedback in adaptive optical systems. Potential applications include live optimization of chirped-pulse amplification systems, stabilization of frequency combs in optical clocks, and dynamic pulse shaping for nonlinear microscopy or laser processing.[11,53–56]

### 4.3 Extensibility to Other Spectrogram-Based Techniques

While demonstrated on SHG-FROG, the architecture generalizes to other ultrafast characterization methods such as GRENOUILLE, d-scan, SPIDER, and transient absorption spectroscopy.[57–60] The generative diffusion approach is particularly suited to any setting where measurements are sparse, noisy, or incomplete—a common scenario in biomedical imaging, atmospheric sensing, and space-based optics.[5,8,9] More broadly, the same "sparse spectrogram inversion" logic for SHG-FROG extends beyond ultrafast pulses to the wider spectrograph ecosystem highlighted in the introduction: snapshot/compressive hyperspectral imagers and integrated/reconstructive spectrometers deliberately acquire coded or reduced measurements (to alleviate acquisition burden) and rely on computational priors to recover full spectral content.[7–9,61]

### 4.4 Societal and Industrial Impact

By making ultrafast pulse characterization more accessible, reliable, and fast, this technology can accelerate innovation across sectors: enabling more efficient solar cell fabrication via ultrafast laser texturing, improving the resolution of multiphoton microscopes for early disease detection, and enhancing the stability of high-power laser systems used in clean energy research. At the same time, by explicitly treating sparsity as a solvable first-class condition rather than a failure mode, the framework connects naturally to broader advanced-photonics priorities in computational imaging and inverse problems, where physics-guided learning and generative priors are increasingly used to reconstruct high-fidelity signals from limited measurements.[62,63]

We have presented a novel generative diffusion framework capable of reconstructing ultrashort pulses from partial FROG traces with high fidelity. This model outperforms both traditional deep learning approaches, such as CNNs, as well as the current state of the art (Seq2Seq) in this domain. CNNs are great at exploiting local patterns in input data to make predictions; however, they struggle to model long-term spectro-temporal relationships when the input is aggressively downsampled. Sequential networks augmented with attention (like Seq2Seq) are better at modeling such dependencies but produce biased estimates on incomplete inputs. These limitations are corroborated by the performance of these models on the unseen test set and the resulting saliency maps. The diffusion model addresses these limitations by denoising a noisy latent vector to obtain the reconstructions with the FROGs acting as an additional mode of context rather than the sole input.

This generative diffusion framework moves ultrafast pulse retrieval beyond incremental algorithmic improvement and toward a new paradigm of accessible, intelligent metrology. By enabling accurate reconstruction from heavily undersampled FROG traces, the model directly addresses the hardware bottlenecks that have historically confined high-fidelity pulse characterization to specialized laboratories. In doing so, it bridges the long-standing gap between research-grade instrumentation and the practical constraints of educational, industrial, and field-deployed photonics systems.

Beyond outperforming existing computational approaches, our framework establishes a pathway toward real-time, high-fidelity diagnostics in environments where temporal or spectral resolution is fundamentally limited. Experimental platforms that cannot support dense scans or premium detectors can still recover full pulse information from sparse measurements, allowing university teaching labs to achieve state-of-the-art characterization with affordable equipment and enabling industrial systems to deploy online feedback and adaptive control during live operation. This capability aligns naturally with emerging efforts in AI-driven laser optimization, digital twins, and autonomous photonic systems, where fast, reliable pulse inference is a prerequisite for closed-loop stability and performance tuning.[56,64–66]

More broadly, this work positions generative AI as an enabling layer for next-generation ultrafast infrastructure. Rather than treating incomplete measurements as a failure mode, our approach treats sparsity as a generative prior, converting limited observations into physically consistent pulse reconstructions. This shift reframes pulse metrology as an inference problem that can scale with modern machine learning, paving the way for portable, affordable, and self-calibrating diagnostic tools. As ultrafast technologies continue to migrate from controlled laboratory settings into high-throughput, resource-constrained, and in situ environments, generative retrieval frameworks such as the one presented here may serve as a foundational component of intelligent photonic systems.

*Acknowledgments*

This work has been supported by the following grants: DOE DE-FOA-0002859, DOE DE-AC02-76SF00515, NSF 2436343, ONR N00014-24-1-2038